*Genetic and population analysis*

# CloudSTRUCTURE: infer population STRUCTURE on the cloud

Liya Wang[1,*], and Doreen H. Ware[1,2,*]

[1]Cold Spring Harbor Laboratory, 1 Bungtown Road, Cold Spring Harbor, NY 11724, USA
[2]USDA-ARS NAA Plant, Soil & Nutrition Laboratory Research Unit, Ithaca, NY 14853, USA

Associate Editor: XXXX

**ABSTRACT**
**Summary:** We present CloudSTRUCTURE, an application for running parallel analyses with the population genetics program STRUCTURE. The HPC ready application, powered by iPlant cyberinfrastructure, provides a fast (by parallelization) and convenient (through a user friendly GUI) way to calculate likelihood values across multiple values of K (number of genetic groups) and numbers of iterations. The results are automatically summarized for easier determination of the K value that best fit the data. In addition, CloudSTRUCTURE will reformat STRUCTURE output for use in downstream programs, such as TASSEL for association analysis with population structure effects stratified.
**Availability:** https://de.iplantc.org/de/
**Contact:** Liya.Wang@cshl.edu, Ware@cshl.edu

## 1 INTRODUCTION

Population structure, or population stratification, is the presence of a systematic difference in allele frequencies between subpopulations due to nonrandom mating, often, as a result of physical separation followed by genetic drift of allele frequencies in each group. For GWAS, failure to account population structure can lead to a spurious associations representing the underlying structure of the population instead of a real disease/trait associated locus. The program STRUCTURE (Pritchard, Stephens et al. 2000) was proposed to estimate and control for population structure with genetic information such as SNPs, and quickly became one of the most widely used programs by population geneticists to assess the level of genetic stratification.

STRUCTURE is a model-based clustering program that relies on multiple MCMC simulations and Gibbs re-sampling for parameter estimation. Therefore, it can be very time consuming for data sets with large numbers of individuals and loci, both of which are increasing quickly with the advances of sequencing technology. The computation time also grows dramatically with the number of genetic clusters (K) to test and the number of replications for each K. Testing multiple K values is required to find the K value best fitting the data based on the output likelihood. Given that the computation for each K value testing is completely independent, it is straightforward to speed up the analysis by distributing the testing to parallel computing cores.

The present application CloudSTRUCTURE, built on top of the iPlant Cyber-infrastructure (Goff, Vaughn et al. 2011), provides a user friendly GUI for customizing STRUCURE parameters, and supports seamless data transfer and parallel analysis on Texas Advanced Computing Center (TACC) clusters. In addition, the output files are automatically converted for the downstream structured association method - TASSEL (Bradbury, Zhang et al. 2007).

## 2 IMPLEMENTATION AND FUNCTIONALITY

CloudSTRUCTURE is built with the command-line version of the STRUCTURE program, which supports limited command-line flags by taking most of the parameters from two text files (usually named as mainparams and extraparams). The mainparams file contains parameters specifying the particular types of data present in the input file, the K value to test, length of burn-in period for minimizing the effect of the starting parameter configuration, and the number of MCMC replications (reps) after burn-in for accurate parameter estimation. The extraparams file contains parameters for model refinement, output options, priors, and miscellaneous options.

The major components of the implementation here include (1) a Graphical User Interface (GUI) built into iPlant Discovery Environment for end users to specify input-output files and parameters presented in the mainparams file, (2) a shell script for generating the mainparams file from the user specified parameters on the fly, preparing a temporary file with a list of commands for parallel analysis with TACC's parametric launcher, and summarizing the results for all tested K values, (3) a JSON (JavaScript Object Notation, a lightweight data-interchange format) file specifying the location of the shell script and the STRUCTURE program, all involved TACC modules, the specific clusters, and properties and ontologies of all parameters.

### 2.1 GUI

As the front-end version of STRUCTURE, the GUI of CloudSTRUCTURE provides convenient specifications for input-output files, major parameters, and estimated maximum runtime for efficiently managing computer clusters. Specifically, parameters in the mainparams file are grouped into two categories, Options and File Information. Under the Options category, in addition to the length of burn-in period and number of reps after burn-in, two more parameters, the maximum K to test and the number of replications for each K can be specified. The multiplication of these two parameters gives the total number of computation processes that can run in parallel. Details of the genotype/marker file (e.g., the number of loci and individuals, ploidy, representation of missing SNP, etc.) are specified under the File Information category.

[*]To whom correspondence should be addressed.

For missing SNPs, the user can choose to impute them with another parallelized DE tool that is built on top of the NPUTE method (Roberts, McMillan et al. 2007). The parallelized version supports not only imputing the missing SNPs but also counting the number of loci and individuals for feeding into CloudSTRUCTURE.

## 2.2 Preprocessing, parallelization and post-processing

The shell script retrieves all STRUCTURE parameters specified in the GUI to create the mainparams file at the run time. If not supplied by the end users, the extraparams file will also be generated on the fly with defaults. The maximum K values to test and the number of replications for each K are used by two nested loops to generate a temporary run file, in which each line represents one run of STRUCTURE with a specified K (from 1 to the maximum number of K with the specified number of replicates). The run file is then used with TACC's parametric launcher module to launch jobs on the parallel cores. Given the specific data set, the length of burn-in periods, and the number of MCMC reps after burn-in, the total computation time is reduced through parallelization, only determined by the run with the largest K value.

By default, STRUCTURE takes a random number seed from the system clock for the MCMC re-sampling. Therefore, the parametric launcher will take the identical seeds for all processes. Due to the property of MCMC, all replications with the same K value will generate identical outputs. To overcome this, the shell script uses the "-D" command-line flag to pass a new random number seed for each run. The random number seed is generated by calling the internal Bash function $RANDOM (that generates a signed 16-bit integer between 0 and 32767) twice as equation (1) for maximizing the randomness.

$$randomSeed = 10000*\$RANDOM+\$RANDOM \quad (1)$$

To overwrite the default random seed, the RANDOMIZE parameter in the extraparams file also needs to be defined as zero (default in the shell script). In cases where the user wants to supply their own extraparams file (e.g., to model with uncorrelated allele frequency), the shell script will check the provided extraparams file and re-define RANDOMIZE variable as zeros. Such randomization ensures that each replication starts with a different configuration, and the consistency among the outputs of all replications can be used to evaluate whether the specified burn-in length and MCMC reps are sufficient to reach convergence. Since increasing any one of these two will significantly increase the computation time, the design here allows user to speed up data analysis by starting with smaller values and relying on checking the consistency among replications for the necessity of additional runs with larger values.

After completion, all results are summarized, and the mean likelihood for each K value are provided for determining the optimum number of genetic groups. The output files associated with each K and replication are also converted for directly feeding into TASSEL for structured association. Users can also feed the output summary file to post-hoc methods, e.g., STRUCTURE HARVESTER (Earl and Vonholdt 2012), for additional analysis. On the system side, all input-output files will be copied back from the parallel cores to the iPlant Data Store, a customized data storage and management system built on top of the integrated Rule-Oriented Data System (iRODS.org).

## 2.3 JSON

The JSON file is used to specify the location of the shell script and the STRUCTURE command-line program in the iPlant Data Store, the TACC cluster for the parallel analysis, modules needed for the analysis, and the property/ontology of the parameters that will in turn generate the GUI. It is the key component describing the application within the iPlant Cyber-infrastructure.

## 3 DISCUSSION

To use CloudSTRUCTURE and all other DE applications, users will need to get an iPlant account and upload data to the iPlant Data Store. Alternatively, for users who are familiar with R and have a manageable computer cluster or multi-core machine, they can install the recently proposed parallelStructure R package (Besnier and Glover 2013) along with the parallelization packages. The advantages of using CloudSTRUCTURE include but are not limited to, a centralized system for data management and analysis on the cloud, convenience for connecting to upstream and downstream applications (e.g., NPUTE and TASSEL), and building automatic workflows with other pre-integrated applications (Wang, Ware et al. 2014).

Combining NPUTE, CloudSTRUCTURE, and TASSEL together, an online tutorial (http://tinyurl.com/m2tkuml) has also been developed for structured association analysis using the Sorghum re-sequencing data as an example.

## ACKNOWLEDGEMENTS

The authors would like to thank the entire iPlant team for developing DE, data store, API, and all underlying CI components. LW would also like to thank Drs. Uwe Hilgert and Christos Noutsos for polishing the tutorial, and thank many users, especially Drs. Elisa Mihovilovich and Alexander D. Twyford for validation and suggestions.

*Funding*: The work is supported by the National Science Foundation Plant Cyber-infrastructure Program (#DBI-0735191).

## REFERENCES

Besnier, F. and K. A. Glover (2013). "ParallelStructure: a R package to distribute parallel runs of the population genetics program STRUCTURE on multi-core computers." PLoS One **8**(7): e70651.
Bradbury, P. J., et al. (2007). "TASSEL: software for association mapping of complex traits in diverse samples." Bioinformatics **23**(19): 2633-2635.
Earl, D. A. and B. M. Vonholdt (2012). "STRUCTURE HARVESTER: a website and program for visualizing STRUCTURE output and implementing the Evanno method." Conservation Genetics Resources **4**(2): 359-361.
Goff, S. A., et al. (2011). "The iPlant Collaborative: Cyberinfrastructure for Plant Biology." Front Plant Sci **2**: 34.
Pritchard, J. K., et al. (2000). "Inference of population structure using multilocus genotype data." Genetics **155**(2): 945-959.
Roberts, A., et al. (2007). "Inferring missing genotypes in large SNP panels using fast nearest-neighbor searches over sliding windows." Bioinformatics **23**(13): I401-I407.
Wang, L., et al. (2014). "A genome-wide association study platform built on iPlant cyber-infrastructure." Concurrency and Computation: Practice and Experience: n/a-n/a.